\documentclass[
aps,superscriptaddress,
twocolumn,prd,
showpacs,preprintnumbers,nofootinbib,
amsmath,amssymb,
aps,
superscriptaddress]{revtex4-1}
\usepackage{Preamble}
\makeatletter
\AtBeginDocument{\let\includegraphics\saved@includegraphics}

\usepackage{cancel}

\newcommand{\cor}[1]
{\textcolor{black}{ #1}}
\usepackage{cancel}

\begin{document}


\title{Identifying Proca-star mergers via consistent ultralight-boson mass estimates across gravitational-wave events}


\author{Ana Lorenzo-Medina}
    \email{ana.lorenzo.medina@rai.usc.es}
    \affiliation{Instituto Galego de F\'{i}sica de Altas Enerx\'{i}as, Universidade de
Santiago de Compostela, 15782 Santiago de Compostela, Galicia, Spain}

\author{Juan Calder\'on~Bustillo}
    \email{juan.calderon.bustillo@gmail.com}
    \affiliation{Instituto Galego de F\'{i}sica de Altas Enerx\'{i}as, Universidade de
Santiago de Compostela, 15782 Santiago de Compostela, Galicia, Spain}
    \affiliation{Department of Physics, The Chinese University of Hong Kong, Shatin, N.T., Hong Kong}

\author{Samson H. W. Leong}
    \email{samson.leong@link.cuhk.edu.hk}
    \affiliation{Department of Physics, The Chinese University of Hong Kong, Shatin, N.T., Hong Kong}

\begin{abstract}

While black-hole and neutron-star mergers are the most plausible sources of current gravitational-wave observations, mergers of exotic compact objects may mimic these signals. Proca stars -- Bose-Einstein condensates of complex vector ultralight bosons -- have gained significant attention for their potential to replicate certain gravitational-wave events while yielding consistent estimates of the boson mass $\mu_{B}$ forming the stars. Using a mixture model within a Bayesian framework, we demonstrate that consistent boson-mass estimates across events can be exploited to obtain conclusive evidence for the existence \cor{of} a number $n$ of Proca-star families characterized by respective boson masses $\mu_{B}^{i}$, even if no individual event can be conclusively identified as such. Our method provides posterior distributions for $n$ and $\mu_{B}^{i}$. Applying this framework to the high-mass events GW190521, GW190426\_190642 \cor{and} GW200220\_061928, we obtain a Bayes Factor ${\cal{B}}^{n=0}_{n=1}=2$ against the Proca-star hypothesis, primarily rooted in the limitation of current Proca-star merger \cor{signal models} to intrinsically weak head-on cases. We show that conclusive evidence $\log{\cal{B}}^{n=1}_{n=0} \geq 5$ could be achieved after 5 to 9 observations of similar event sets, at the $90\%$ credible level. Our framework provides a new way to detect exotic compact objects, somewhat using gravitational-wave detectors as particle detectors.
\end{abstract}

\maketitle

\section{Introduction}

The gravitational-wave (GW) detectors, Advanced LIGO~\cite{AdvancedLIGOREF} and Advanced Virgo~\cite{TheVirgo:2014hva}, now joined by KAGRA~\cite{akutsu2020overview,Somiya:2011np}, have made the observation of compact binary mergers almost routine. During their first three observing runs, these detectors have reported $\mathcal{O}(100)$ such observations that have provided us with unprecedented knowledge on how black holes (BHs) and neutron stars (NSs) form and populate our Universe~\cite{LIGOScientific:2018mvr,abbott2021gwtc2,abbott2021gwtc3,GWTC4_catalog,GWTC2-pop,Populations_GWTC3,Pop_GWTC4}. Moreover, these observations have enabled the first tests of General Relativity in the strong-field regime~\cite{TGR_GWTC2,GWTC3-TGR} and qualitatively new studies of the Universe at  large scales~\cite{H0_nature_lvk,Abbott2021_gwtc12_ho,Abbott2023_gwtc3_h0}.

While BHs and NSs stand as the most plausible and anticipated GW sources, there exist further theoretical proposals of exotic compact objects (ECOs) that can mimic their GW emission~\cite{compact_objects_nico_miguel,Pani2010,Cardoso:2016oxy,Toubiana2021,brito2016proca,sanchis2019head,Tamara_Isobel_boson_stars,Asali2020,Psi4_obs_PRD}. A particularly appealing class of ECOs is that of boson stars~\cite{Ruffini:1969qy,Kaup:1968zz}, which are among the simplest and most dynamically robust ECOs one can consider. These consist on Bose-Einstein condensates of ultralight bosons that can clump together to form macroscopic objects of astrophysical size. Among boson stars, Proca-stars~\cite{brito2016proca} formed by complex-vector bosonic fields have recently arisen as extremely appealing candidates~\cite{brito2016proca} due to their particular properties. While scalar bosonic fields can \cor{also} form boson stars, \cor{they} are unstable against non-axisymmetric perturbations and cannot therefore be stable if they have spin, unless self-interactions are considered \cite{Siemonsen2021}. Spinning Proca stars, in contrast, are stable and can therefore survive long enough to undergo mergers that can produce GW emission detectable by GW detectors~\cite{sanchis2017numerical,sanchis2019head}. Furthermore, vector bosons are not only motivated in some extensions of the Standard Model of elementary particles but have also been long considered as plausible candidates to form part of what we know as Dark Matter \cite{Arvanitaki2010,Freitas:2021cfi}. 

\begin{figure}[t!]
\begin{center}
\includegraphics[width=0.495\textwidth]{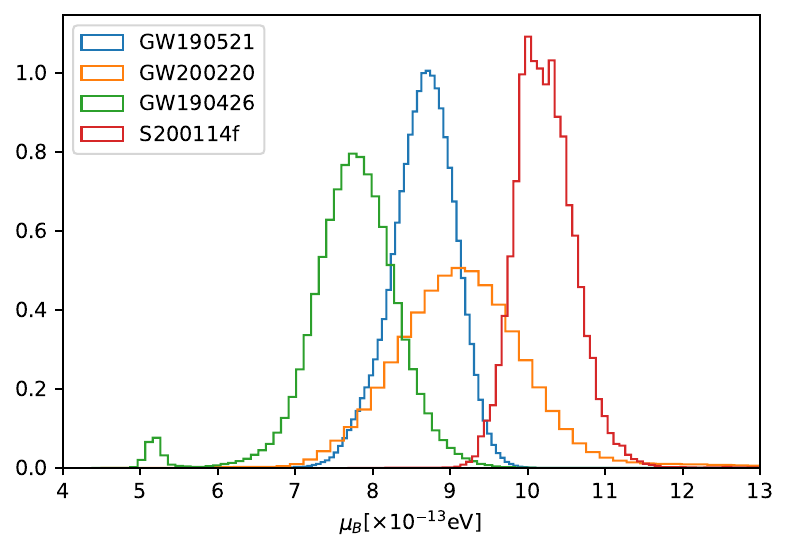}
\caption{\textbf{Boson-mass estimates for the events we consider in this work.} Posterior distributions for the boson mass for the four events we consider, obtained under the \cor{assumption} that these are sourced by head-on Proca-star mergers. Details on the corresponding analysis can be found in \cite{Psi4_obs_PRD}.}
\label{fig:boson_masses}
\end{center}
\end{figure}

Several groups have succeeded to obtain numerical simulations of boson-star mergers, extracting the corresponding GW emission~\cite{sanchis2019head,Nico_phases,Tamara_Isobel_boson_stars}. The latter can then be used to construct \cor{waveform} templates to search for these objects in GW data and in parameter inference tasks~\cite{Proca,Psi4_obs_PRD,Psi4_PRX,Luna2024,Tamara_Isobel_boson_stars}. In particular, although still restricted to the head-on merger case, both numerical simulations of Proca-star mergers (PSMs) and continuous surrogate waveform models have recently been compared to sufficiently short GW signals detected by LIGO and Virgo~\cite{Proca,Psi4_obs_PRD,Luna2024,Psi4_PRX}. These analyses have found that the signal GW190521~\cite{GW190521D} is indeed consistent with a \cor{head-on} Proca-star merger, with this scenario being weakly preferred over the ``vanilla'' BBH one. In addition, a mass $\mu_B \simeq 8.7 \times 10^{-13}$ eV was estimated for the complex ultralight boson building up the stars. A recent follow-up was performed on two extra ``catalogued'' GW events, namely GW190426$\_$190642~\cite{GWTC-2.1_PRD}, GW200220$\_$061928~\cite{GWTC-3_PRX}, which we will refer to as simply \cor{GW200220 and GW190426.} Interestingly, while showing statistical preference for the BBH scenario, these two events yield a boson-mass $\mu_B$ consistent with that \cor{inferred for} GW190521 (see Fig. \ref{fig:boson_masses}). Finally, a fourth GW trigger known as S200114$\_$020818~\cite{O3IMBH_AA} (which we will refer to as S200114) weakly favours the PSM hypothesis, albeit yielding a boson mass inconsistent with GW190521.\\

Notably, all of the above events display component black-hole masses inconsistent with \cor{a} stellar-collapse origin, as these fall within the so-called Pair-instability Supernova (PISN) gap~\cite{Heger:2002by}. In this context, exotic-compact objects arise as a possible alternative explanation to hierarchical formation channels~\cite{Gerosa2021_Fischbach_review} which, while more plausible than ECOs, require restrictive merger configurations that lead to small gravitational recoils \cite{Gonzalez:2006md,Varma_strong_kick,GW190412_kick} that prevent remnant black holes from leaving their host environments~\cite{Arajolvarez2024,Measured_kick_magnitude_GW190814,Mahapatra2024}. Similarly, ECOs shall also offer alternative explanations~\cite{Mourelle2024} to low-mass objects that are slightly above the expected neutron-star mass and below that of stellar-born black holes~\cite{virgo2020gw190814,Abac2024,2503.17872}. 

Both the gain in sensitivity of GW detectors towards their fourth observing run and the future existence of next-generation detectors like Cosmic Explorer~\cite{US_CE,CE_Horizon_study}, Einstein Telescope~\cite{Hild2011_ET,Punturo2010,ET_Science} or LISA~\cite{LISA} make it reasonable to anticipate an important increase in the number of observed events similar to those discussed above. This, together with the development of improved templates for PSMs, will provide a great arena to probe the existence of potential sub-populations of PSMs mixed with ``vanilla''  BBHs. The identification of elements from such Proca-star populations will involve, at least, the comparison of individual events to both PSM and BBH models through a Bayesian framework to perform model selection. This will return posterior distributions on the individual source parameters including those for boson mass $\mu_B$. Model selection between BBH and PSM models across many observations may conclude that no individual event yields decisive evidence favouring the PSM model. On the one hand, this is may be due to the fact that current GW searches precisely target BBH signals~\cite{Usman:2015kfa,Messick:2016aqy,Chandra2022_IMBHHMsearch}, so that any \cor{detected GW} should be sufficiently similar to a BBH. On the other hand, it is known that in some cases BBHs and PSMs can emit extremely similar signals~\cite{Proca,Psi4_obs_PRD}, making their discrimination intrinsically challenging. In this work we will show, however, that consistent estimations of parameters such as the boson mass ($\mu_B$) across PSM candidates can be exploited to conclusively identify specific sub-populations of PSMs characterized by distinct boson masses. Finally, even in a hypothetical situation where several conclusive PSM candidates already exist, our formalism can be leveraged to decide the number of ultralight bosons responsible for such candidates.\\ 

The rest of this work is organised as follows. In section I, we introduce our working tool: a mixture model considering a mixed observation set consisting of BBHs mixed with $n$ families of PSM mergers respectively characterised a boson mass $\mu_{B}^{i}$ -- with $i \in \{1,n\}$ -- and a parameter $\zeta_{i}$  denoting the fraction of observations belonging to such family. Second, we apply this formalism to two different sets of existing GW observations that respectively exclude and include the trigger S200114. Performing analyses allowing for different numbers $n$ of ultralight bosons, we obtain Bayesian evidences for each value of $n$ and posterior distributions for the parameters $\mu_{B}^{i}$, $\zeta_{i}$. We also discuss the impact of prior choices. Finally, in Section III, we showcase the power of this formalism to detect PSM sub-populations on mock data sets consisting on several instances of event sets similar to the ones analysed in Section II. We show that the observation of 5-9 such sets would \cor{enable} a conclusive identification of a Proca-star merger sub-population while no evidence would be found if mass consistencies \cor{across} events are ignored.

\section{Method} 

\subsection{A Proca-star counting experiment}

Consider a dataset consisting on $N$ GW observations \cor{$d=\{d_i\}$ with $i \in \{1, \ldots, N\}$.} Next, consider a mixture model where a fraction $\zeta$ of the events correspond \cor{to} PSMs while a fraction $1-\zeta$ correspond \cor{to} BBHs. The likelihood for the fraction parameter $\zeta$ can be expressed as~\cite{Callister2022_mixedmodel}:

\begin{equation}
\begin{aligned}
    & p(\{d_i\}|\zeta) \\ & = \prod_{i=1}^{N} \bigg{[}p(d_i|\text{PSM})\zeta + p(d_i|\text{BBH})(1- \zeta)\bigg{]} \\ & \propto \prod_{i=1}^{N} \bigg{[} {\cal{B}}^{\text{PSM}}_{\text{BBH},i}\zeta + (1-\zeta)\bigg{]}.
\end{aligned}
\label{eq:popnoboson}
\end{equation}

Above, the terms $p(d_i|\text{PSM})$ and $p(d_i|\text{BBH})$ denote respectively the probability of the data $d_i$ of the $i-th$ GW event given the PSM and BBH models against the noise hypothesis -- also known as Bayesian evidence -- while the term ${\cal{B}}^{\text{PSM}}_{\text{BBH},i}$ denotes their ratio. In general, given a signal model $A$ depending on parameters $\vec\theta$, the Bayesian evidence is \cor{defined as} 
\begin{equation}
    p_A(d) \equiv p_{\rm marg}(d|\vec\theta) = \int p(\vec\theta) p(d|\vec\theta) \,\dd \vec \theta, 
 \end{equation}
with $p(\vec\theta)$ denoting the prior probability for the parameters $\vec\theta$ and $p(d|\vec\theta)$ denoting their likelihood given the data $d$.

\subsection{Counting Proca-stars under a common boson mass}

Under the BBH scenario, the parameters $\vec\theta$ denote the masses and three-dimensional spins \cor{of the two component BHs} in addition to the extrinsic source parameters: the two sets of angles respectively parametrising the source orientation and sky-location, the luminosity distance, the coalescence time and the signal polarization. For the case of PSMs, we work with \cor{results obtained in \cite{Psi4_obs_PRD} under the restriction to} head-on merger configurations where both spins are aligned between them and with the final spin. In addition, the star spin is linked to the mass-normalised oscillation frequency \cor{of the boson field} $\omega/\mu_B$. \cor{Combined with the mass of the star, $\omega/\mu_B$ determines the ultralight boson mass $\mu_B$.} \cor{We note that the restriction of our analysis to head-on, aligned-spin configurations is caused by existing limitations in the obtention of reliable initial data to be used in numerical simulations of more generic configurations. This has to main effects. First, it will naturally limit the ability of PSM simulations to fit existing GW data, leading to small evidences for such scenario. Second, it will also potentially lead to over-constrained posterior distributions on the corresponding boson mass.}

If Proca stars exist, it seems plausible that these may come in the form of a discrete number of sub-families, \cor{each characterised by a value of the boson} mass \cite{Arvanitaki2010,Freitas:2021cfi}. This is, each family would be sourced by a \cor{common ultralight boson with mass $\mu_B^{i}$ with $i \in \{1,...,n\}$ and $n$ denoting the number of permitted ultralight bosons}. To start with, we will discuss the $n=1$ case, \cor{where we consider} a single PSM family characterised by a boson mass $\mu_B$. Such model can be implemented by simply requiring that all analysed events share a common boson mass $\mu_B$, as opposed to freely estimating such \cor{parameter} for each individual event. Such condition has an important consequence in terms of model selection: it reduces \cor{the} number of sampled boson-mass parameters from $N$ (the number of considered events) to $n=1$, reducing the complexity of the model and, therefore, the incurred Occam Penalty. This will cause an increase of its evidence with respect to the BBH model \textit{if all events point to the same boson mass}. On the contrary, if the individual events point to different masses, the model will not be able to fit the data, making the evidence for the PSM model decrease.\\

The joint likelihood for the fraction $\zeta$ of PSMs in our dataset and the common boson mass $\mu_B$ is given by

\begin{equation}
\begin{aligned}
    & p(\{d_i\}|\zeta,\mu_B) \\ & = \prod_{i=1}^{N} \bigg{[}p(d_i|\text{PSM}(\mu_B))\zeta + p(d_i|\text{BBH})(1- \zeta)\bigg{]} \\ & \propto \prod_{i=1}^{N} \bigg{[} {\cal{B}}^{\text{PSM}}_{\text{BBH},i}(\mu_B)\zeta + (1-\zeta)\bigg{]},
\end{aligned}
\end{equation}

\cor{Above, the quantity $p(d_i|\text{PSM}(\mu_B))$ denotes the Bayesian evidence for the PSM model assuming a prior probability for the boson mass defined by a delta placed \cor{at} a fixed value of $\mu_B$.} This can be simply obtained from existing analyses making use of wider priors on $\mu_B$ through the Savage-Dickey density ratio as

\begin{equation}
   p(d_i|\text{PSM}(\mu_B)) = p(d_i|\text{PSM}) \frac{p(\mu_B | d_i)}{p(\mu_B)} ,
\label{eq:overlap}
\end{equation}

where $p(\mu)$ denotes the prior \cor{probability} on the boson mass $\mu_B$ imposed in such existing analyses and $p(\mu_B|d_i)$ denotes the corresponding posterior probability.

\subsection{A boson-counting experiment: \\ allowing for several boson masses}

We now generalise the above formalism to \cor{a case where we allow for} $n$ Proca-star families, \cor{each} characterised by \cor{a value of the boson mass} $\mu_{B}^{j}$ with $j \in [1,n]$. In this case, the \cor{joint} likelihood for the $n$ boson masses $\mu_B^j$ and the corresponding fractions  $\zeta_j$ \cor{of the $N$ analysed events belonging to each family within the observed data} is given by

\begin{widetext}
\begin{equation}
\begin{aligned}
    & p(\{d_i\}|\{(\mu_B^j,\zeta_j)\}) \\ 
    &= \prod_{i=1}^{N} \left[ \sum_{j=1}^{n}  p(d_i|\text{PSM}(\mu_B^j))\zeta_j + p(d_i|\text{BBH})\qty(1-\sum_{j=1}^{n}  \zeta_j)\right] 
    & \propto \prod_{i=1}^{N} \qty[ \sum_{j=1}^{n} {\cal{B}}^{\text{PSM}}_{\text{BBH},i}(\mu_B^j)\zeta_j + \qty(1-\sum_{j=1}^{n} \zeta_j)].
\end{aligned}
\label{eq:pop}
\end{equation}
\end{widetext}

\begin{table*}
\centering
\begin{center}

\renewcommand{\arraystretch}{1.5}
\begin{tabular}{c|cc|cc|cc|cc}
\rule{0pt}{3ex}%
Event & \multicolumn{2}{c}{GW190521}  &  \multicolumn{2}{c}{GW200220} & \multicolumn{2}{c}{GW190426} & \multicolumn{2}{c}{S200114}  \\
\hline
    &  V & D &  V & D &  V & D &  V & D  \\

Black hole merger & 89.6  & 89.7  & 17.4 & 17.4 & 37.9 & 38.2    & 69.1 & 71.0 \\
Proca star merger &  90.7  & 93.4  & 13.4  & 15.5  & 29.5  & 32.4 &  71.1 & 76.3 \\

$\log\cal{B}^{\text{PSM}}_{\text{BBH}}$ & 1.1  & 3.7  &  -4.0  & -1.9 &  -8.4  & -5.8 & 2.0 & 5.3 \\
$\cal{B}^{\text{PSM}}_{\text{BBH}}$ & 3.0  & 40.5  &   0.02  & 0.15 &  $2\times 10 ^{-4}$  & $3\times 10 ^{-3}$ & 7.2 & 200.3 \\

\end{tabular}
\caption{\textbf{Bayesian evidences for the black-hole and Proca-star merger hypotheses for the events considered in this work.} The two columns for each event denote the natural log Bayes factors (signal ${\it v.s.}$ noise) obtained using either a standard prior uniform in co-moving volume (V) or a prior uniform in luminosity distance (D). The last two rows denote respectively the corresponding relative log Bayes factors and their exponentials.}
\label{tab:logb1}
\end{center}
\end{table*}

Finally, the Bayesian evidence for a model \cor{allowing for} a number of bosons $n$ is just given by

\begin{equation}
     p(\{d_i\}|n)  = \int p(\{d_i\}|\{\zeta_j,\mu_B^{j}\})  \prod_{j = 1}^n p(\zeta_j, \mu_B^{j}) \, \dd\zeta_j \dd\mu_B^j\ ,
\end{equation}


where $i \in [1, N]$.\\ 

With this, we can construct the posterior distribution on the number of bosons $n$ as

\begin{equation}
\begin{aligned}
     p(n|\{d_i\}) \propto p(\{d_i\}|n) \pi(n),
\end{aligned}
\end{equation}

where we will always consider an uniform prior probability $\pi(n)$ on $n$. \cor{Finally, we note that the likelihood for the case $n=0$, where all events are considered to be BBHs, is simply given by setting $\zeta_i=0$ in Eq.~\eqref{eq:pop}, which is equivalent to setting $\zeta=0$ in Eq.~\eqref{eq:popnoboson}.}

\subsection{Parameter sampling}

We impose uniform priors on the boson masses with $\mu_B^i \in [2,15]\times10^{-13}$\,eV. \cor{This range safely contains the full posterior distributions for $\mu_B$ individually inferred for the four events we study\footnote{We also note that setting $\mu_B=0$ reduces the Proca equation to Maxwell’s equations, corresponding to the massless photon case. In this limit the field supports only long-range radiative solutions, which prevents the existence of localized, gravitationally bound configurations that could form compact objects.}.} \cor{In order to label the different bosons and avoid double-counting, we choose to impose} $\mu^{i+1}_B \leq \mu^{i}_B$. Importantly, for computational-cost reasons, we do not sample over a continuous range of values $\mu^i_B$. Instead, \cor{we sample} over a discrete grid with step $\delta \mu_B = 1\times 10^{-15}$ eV. Physically, $\delta \mu_B$ can be interpreted as the maximum resolution in the boson mass we allow for, which is much smaller than the standard deviation \cor{of the individual posteriors for each event that we take as input}. We also set a flat prior on the respective fractions $\zeta_i \in [0,1]$, imposing $\sum_i \zeta_i \leq 1$. We sample the parameter space using the nested sampler Dynesty \cite{Dynesty} with 4096 live points. 

\subsection{Inputs for our analysis}

We take as \cor{input data} the Bayesian evidences \cor{$p(d_i|{\rm BBH})$ and $p(d_i|{\rm PSM})$} for the BBH and PSM models for the four events we consider reported in~\cite{Psi4_obs_PRD}, \cor{together with the corresponding posterior distributions $p(\mu_B|d_i)$ for the boson mass obtained under the PSM model}. These are respectively shown in Fig.~\ref{fig:boson_masses} and Table~\ref{tab:logb1}. In the mentioned study, the authors performed Bayesian parameter inference and model selection on the events we consider, both under the BBH and PSM scenarios. For the BBH case, the data was analyzed with the state-of-the art waveform model for BBHs \texttt{NRSur7dq4}~\cite{NRSur7dq4}. This model includes waveform multipoles up to $\ell = 4$. The model is directly calibrated to numerical simulations of generically spinning BBHs~\cite{SXSCatalog} with mass ratios $q\in [1,4]$ and spin-magnitudes $a_i \in [0,0.8]$ but can be extrapolated to $q=6$ and extremal spins. For the PSM case, the authors employed a family of 769 numerical simulations of unequal-mass head-on mergers of Proca stars, with the corresponding field frequencies ranging in $\omega/\mu_B \in [0.80,0.93]$ and imposing a flat prior on these parameters. In these configurations, the stars were released at rest at a separation distance of $20\,M$, expressed in geometric units. \cor{For a visualization of the GW detector data for the four events we study, together with the corresponding best-fitting BBH and PSM waveforms, we refer the reader to Figs. 1-4 in~\cite{Psi4_obs_PRD}.}

Interestingly, the above analysis also considered two different distance priors. First, a ``regular'' and physically realistic prior uniform in co-moving volume with Hubble parameter H$_0=67.74$\,km\,s$^{-1}$\,Mpc$^{-1}$. Second, a prior uniform in luminosity distance. The motivation for using the latter prior relies on the fact that head-on merger simulations considered for our input analyses are much less luminous than quasi-circular ones. This leads to a much weaker GW emission that makes the BBHs, which can produced the same signal from a much further distance. \cor{This makes BBHs to be significantly favoured} by the realistic distance prior. On the one hand, the usage of the uniform distance prior allows us to check the \cor{impact of this effect on model selection}. On the other, assuming that \cor{future numerical simulations of} quasi-circular BBS can also match these GW events, it serves us as a ballpark estimation of how our evidences may be modified if using such a model\footnote{An alternative option to increase the loundess of the source keeping the head-on configuration would be to simply boost the stars, unlike in~\cite{Proca,Psi4_obs_PRD}, where they start at rest.}. \cor{We stress that no meaningful conclusions at the astrophysical or population level can be extracted by using this last prior. Instead, this study allows us to explicitly isolate and understand the impact, in terms of Bayesian model selection, of the intrinsic weakness of the PSM sources to which the GW data was compared.} We will describe results obtained under both distance priors.

\section{Results on real data}

\begin{figure}
\begin{center}
\includegraphics[width=0.49\textwidth]{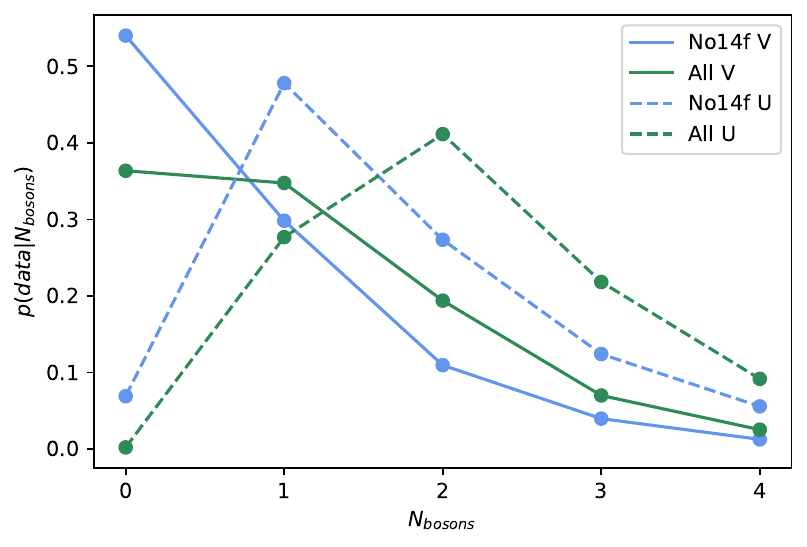}
\caption{\textbf{Posterior probability for the number $N_{\rm bosons}$ of ultra-light bosons given \cor{the Advanced LIGO - Virgo events we consider}.} Solid (dashed) lines ignore (include) the event S200114 in the analysis. Blue and green lines respectively make use of distance priors uniform in co-moving volume and on luminosity distance. \cor{Note that, in order to alleaviate notation, in the main text we denote the variable $N_{\rm bosons}$ simply by $n$.}
}
\label{fig:posterior_n}
\end{center}
\end{figure}

\begin{figure}
\begin{center}
\includegraphics[width=0.49\textwidth]{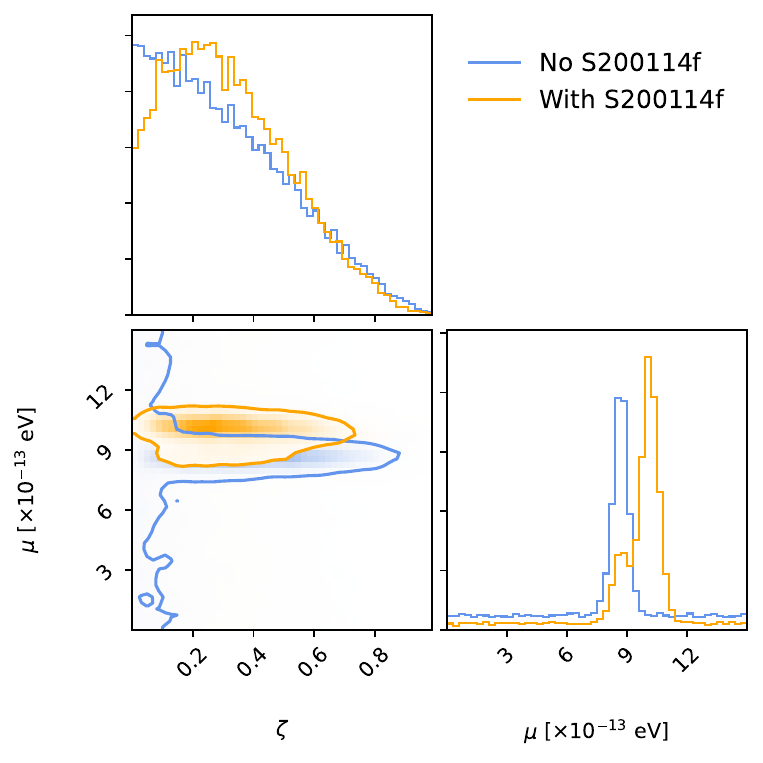}
\caption{\textbf{Posterior distributions for the boson-mass and Proca-star fraction \cor{obtained when we only allow for a single ultralight boson}.} Blue and orange curves and contours denote results \cor{obtained by respectively} including \cor{and} excluding and including the S200114 trigger. The results are obtained assuming a distance prior uniform in co-moving volume. The posterior distribution for the boson mass has non-zero support for all the prior range because the posterior distribution for $\zeta$ is non-null at $\zeta = 0$, in which case all values of $\mu_B$ are allowed.   
}
\label{fig:posterior_mu_zeta_1boson}
\end{center}
\end{figure}

\begin{figure*}[t!]
\begin{center}
\includegraphics[width=0.495\textwidth]{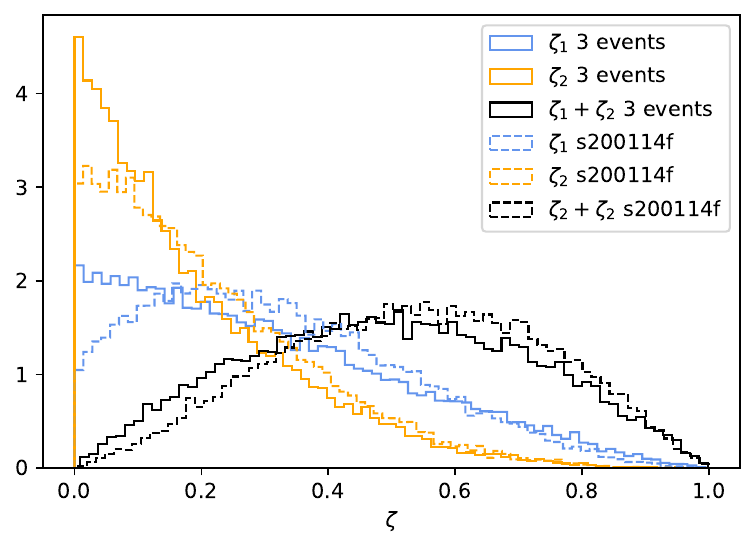}
\includegraphics[width=0.498\textwidth]{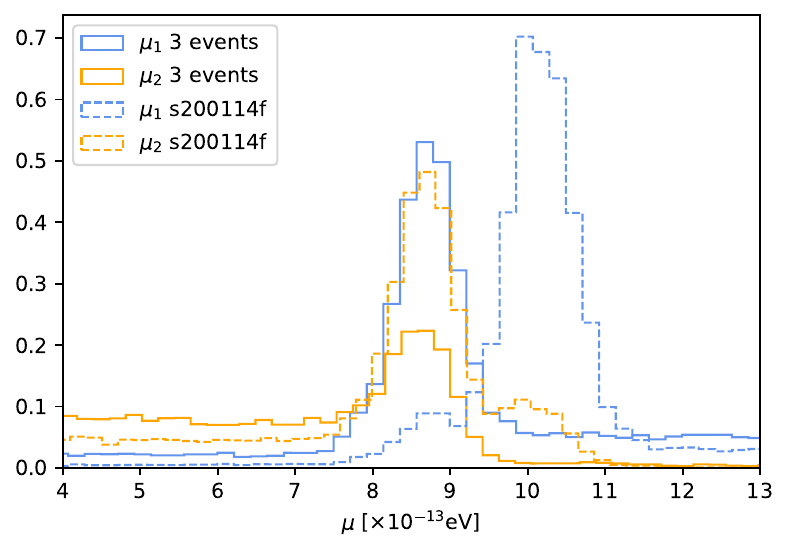}
\caption{\textbf{Posterior distributions for boson-mass (right) and Proca-star fractions (left) for the 2-boson model.} Solid and dashed curves denote results \cor{respectively} excluding and including the S200114 trigger. Blue and orange curves respectively denote the value of the primary and secondary boson-mass $\mu_B^{1,2}$ and \cor{the} corresponding fraction $\zeta_{1,2}$. Black curves denote the posterior distribution for $\zeta_1+\zeta_2$.}
\label{fig:posterior_zetas_mus}
\end{center}
\end{figure*}

Fig.~\ref{fig:posterior_n} shows the posterior probability for the number of ultralight bosons $n$ given our different data sets and priors. Blue and green curves respectively exclude and include the trigger S200114. We start by discussing results \cor{obtained} under the regular distance prior. In this situation, \cor{the} $n=0$ model is preferred \cor{with a Bayes factor} $ {\cal{B}}^{n=0}_{n=1}\simeq 1.6$ with respect to the $n=1$ model. This weak preference \cor{for} $N=0$ is expected, as Table~\ref{tab:logb1} shows that only one of the analyzed events has a slight preference for the PSM scenario while the other two reject such hypothesis rather strongly. In other words, the slight preference of GW190521 for the PSM hypothesis is consistent with a statistical fluctuation once the three events are jointly considered. 

Fig.~\ref{fig:posterior_mu_zeta_1boson} shows, in blue, the two-dimensional $90\%$ credible region for $\mu_B$ and $\zeta$ \cor{under} the $n=1$ model obtained when we exclude S200114, together with the corresponding one-dimensional posterior distributions. We obtain a boson mass $\mu_B=8.64^{+4.84}_{-7.17} \times 10^{-13}\,\rm{eV}$ and a fraction of $\zeta = 0.28^{+0.43}_{-0.25}$ at the $90\%$ credible level. The value of $\mu_B$ is consistent with that obtained from the individual analysis of GW190521 in~\cite{Proca}. Notably, however, the uncertainty is significantly larger due to the fact that $\zeta$ is consistent with $0$, in which case no constraints can be imposed on $\mu_B$.\\

The solid-green line in Fig.~\ref{fig:posterior_n} shows that the inclusion of S200114 does not change our qualitative conclusions. Quantitatively, however, it makes the $n=0$ and $n=1$ become equally preferred. The reason is that, while there \cor{are now} two events favouring the PSM hypothesis, these point to inconsistent boson masses. This makes the $n=1$ model having to decide whether GW190521 or S200114 is \cor{a} better PSM candidate -- as both cannot be at the same time if only one boson mass is allowed -- preventing a larger evidence in favour of such model. The corresponding posterior distributions for $\mu_B$ and $\zeta$ for the $n=1$ case are shown in orange in Fig.~\ref{fig:posterior_mu_zeta_1boson}. The most remarkable difference with respect to the case omitting GW190521 is that, as expected, the inferred boson mass peaks now at the value obtained in the individual analysis of S200114, as this event has a larger preference for the PSM model than GW190521.\\

The right and left panels of Fig.~\ref{fig:posterior_zetas_mus} show respectively our posterior distributions for $\mu^{1,2}_B$ and $\zeta_{1,2}$ for the case $n=2$ where we allow for the existence of two bosons. To give an idea of the ``level of preference'' \cor{for this model {\it v.s.} the $n=0$ one}, we also show the posterior for $\zeta_1 + \zeta_2$. Solid and dashed curves respectively correspond to analyses excluding and including S200114. We highlight two main aspects. First, the posteriors for $\mu^{1,2}_B$ are clearly inconsistent with each other when S200114 is included in the analysis, each being respectively consistent with the mass values of S200114 and GW190521. Second, as shown in Fig. \ref{fig:posterior_n} the evidence for the $n=2$ model is however slightly smaller than for $n=1$. This happens because the increase of the likelihood produced by the fact that each of the two events can now be fit through a different boson mass cannot overcome the increased Occam Penalty paid by \cor{the two extra degrees of freedom of this model}. This result may, at first sight, seem in contradiction with the fact that the posterior for $\zeta_1 + \zeta_2$ clearly peaks away from zero, where it achieves null values. The interpretation of such posteriors for model selection, however, needs to fold in the fact that these are heavily influenced by the corresponding prior, which peaks at $\zeta_1 + \zeta_2 = 1$. In fact, a rough estimation of the Bayes Factor for the $n=2$ {\it v.s.} $\zeta_1 + \zeta_2 = 0$ ({\it i.e.} $n=0$) model via the Savage-Dickey ratio yields values between $1.3$ and $1.7$ favouring $n=0$. This is consistent with the values shown in Fig.~\ref{fig:posterior_n} (green). 

\subsection*{Prior uniform in luminosity distance: impact on model selection}

\cor{We now study the impact of the intrinsic weakness of head-on PSM mergers in the above results by considering a prior uniform in distance that removes such effect from our analysis. While this prior and the corresponding conclusions in terms of parameter inference not physically meaningful, it allows us to understand how results under our regular prior are impacted by the intrinsic weakness of head-on PSMs. We indeed see that this has a tremendous impact in our analysis.} Switching to this prior causes both GW190521 and S200114 to strongly favour the PSM hypothesis. As a consequence, even if S200114 is excluded from the analysis, we obtain ${\cal{B}}^{n=1}_{n=0}\simeq 8$ when exploiting boson-mass coincidences and ${\cal{B}}^{n\neq 0}_{n = 0} \simeq 3$ when ignoring them. If S200114 is included then the $n=2$ model is preferred due to its boson mass being inconsistent with that of GW190521. In particular, we obtain $ {\cal{B}}^{n=2}_{n=0}\simeq 50$ and ${\cal{B}}^{n\neq 0}_{n = 0} \simeq 40$ and $ {\cal{B}}^{n=2}_{n=1}\simeq 1.3$. Needless to say, these results do not imply by any means an actual detection of PSMs. These, however, underscore the potential impact that the production of more realistic simulations of \textit{louder} PSMs configurations -- which would not be penalised by a realistic distance prior -- could have in future studies searching for these objects.

\begin{figure*}[t!]
\begin{center}
\includegraphics[width=0.49\textwidth]{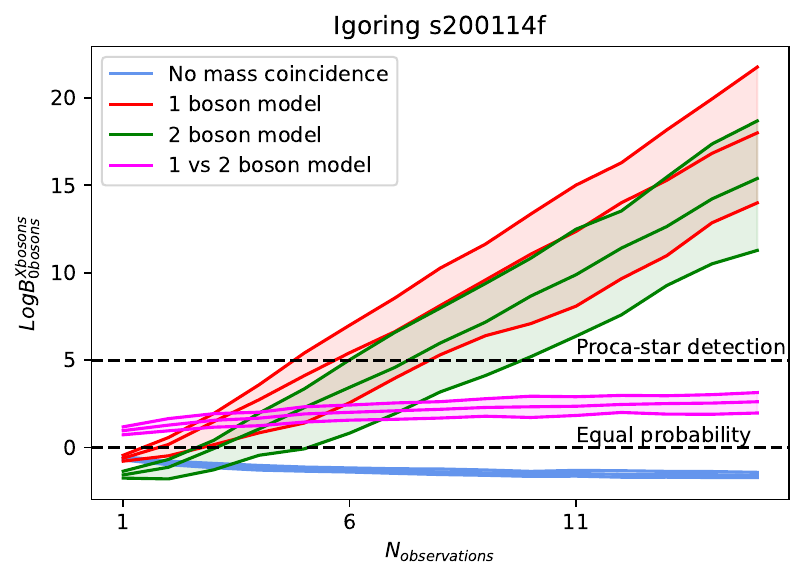}
\includegraphics[width=0.49\textwidth]{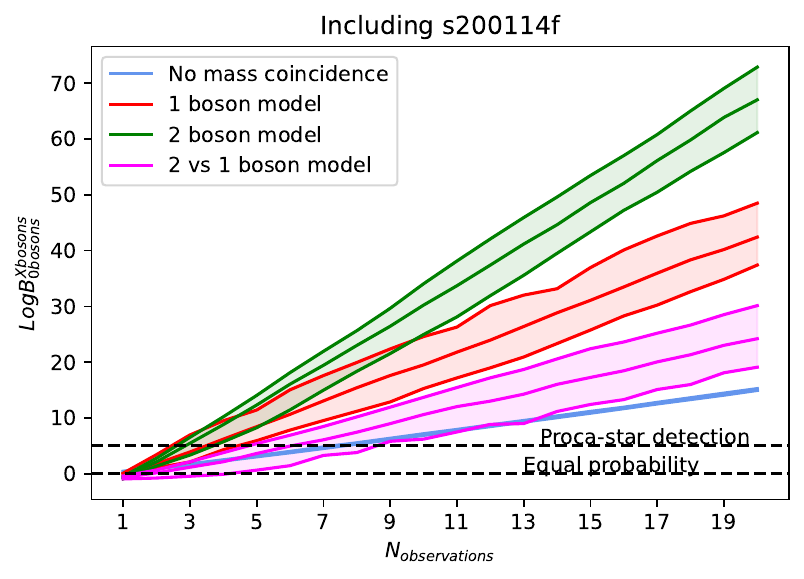}
\caption{\textbf{Accumulated evidence for Proca-star mergers as a function of the number of observations}. Left: $90\%$ credible intervals for the natural log Bayes Factor for the $N_{\rm bosons} = 1$ (red) and $N_{\rm bosons} = 2$ (green) bosons {\it v.s.} $N_{\rm bosons} = 0$ as a function of the number of times that event sets like GW190521+GW190426+GW200220 are observed. The blue contour shows the evidence for non-zero {\it v.s.} zero Proca-star mergers when potential boson-mass coincidences across events are ignored. The magenta contour shows the log Bayes Factor for $N_{\rm bosons} = 1$ {\it v.s.} $N_{\rm bosons} = 2$. Right: Same as in the right panel, including the event S200114 in the event sets. Here, the magenta contour denotes the log Bayes factor for $N_{\rm bosons} = 2$ {\it v.s.} $N_{\rm bosons} = 1$. 
}
\label{fig:forecast_n}
\end{center}
\end{figure*}

\begin{figure*}[t!]
\begin{center}
\includegraphics[width=0.49\textwidth]{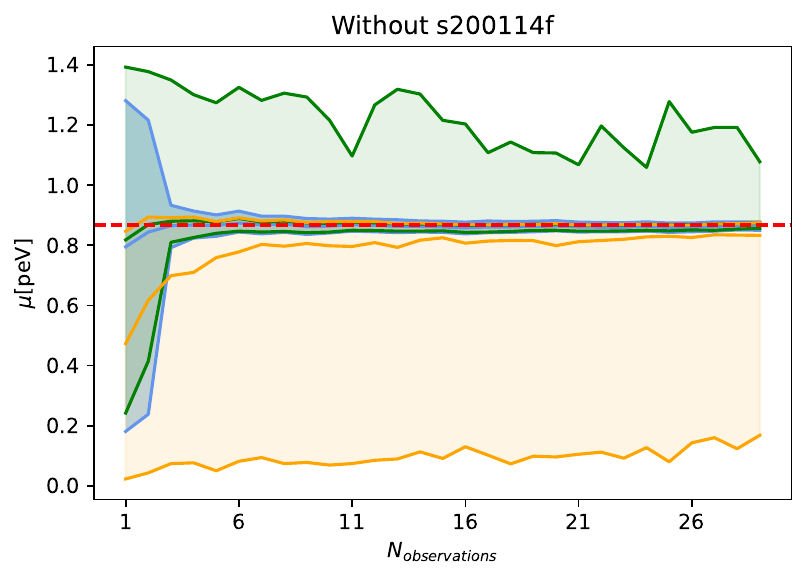}
\includegraphics[width=0.49\textwidth]{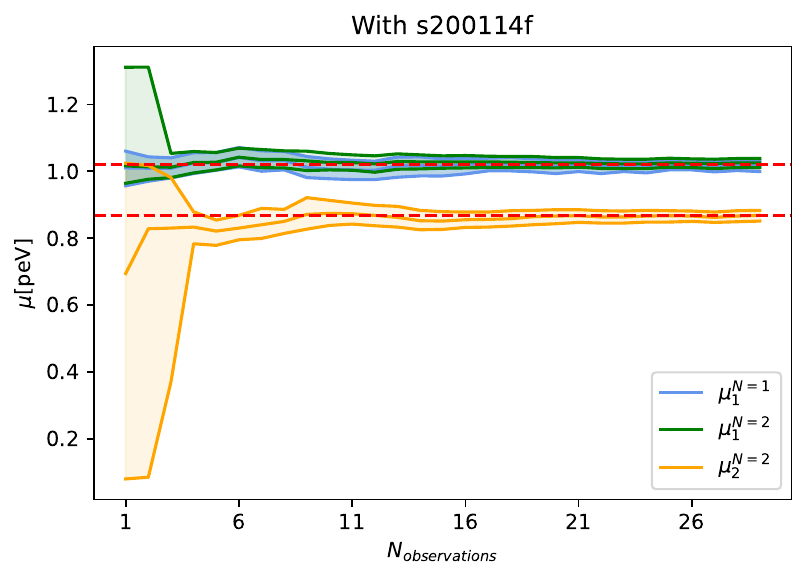}
\caption{\textbf{Evolution of the posterior distribution for the ultralight-boson mass as a function of the number of detected events}. Left: $90\%$ credible intervals as a function of the number $N_{\rm obs}$ of observations of event sets like GW190521+GW190426+GW200220. Right: same as left but adding a S200114-like event to the event set. The red line denotes the median values of the boson mass obtained for GW190521 and S2001124.}
\label{fig:forecast_mass}
\end{center}
\end{figure*}

\section{Detecting Proca-stars in mock data sets}

\cor{In this section}, we investigate the power of our framework to extract evidence for a population of PSMs and to estimate the corresponding boson mass(es) as observations accumulate. \cor{In particular, we want to consider events similar to the ones at hand, which satisfy that a) no event can individually be identified as a PSMs in a conclusive way b) several of these events actually point to similar boson masses.} To do so, we simulate the observation of $N_{\rm obs}$ instances of the \cor{two sets of real events we have analyzed in this work, respectively ignoring and including simulated events similar to S200114f}. Each simulated observation will retain the same relative evidences for the BBH and PSM models as the events we have analyzed, but incorporates statistical fluctuations in the corresponding posterior distribution of the boson-mass. Specifically, we generate such perturbed posteriors $p^{*}(\mu_B)$ by shifting the original posteriors $p(\mu_B)$ by random offsets $\delta_{\mu_B}$, sampled from the distribution $p(\mu_B - \mu^{50}_{B})$, where $\mu^{50}_{B}$ denotes the median of the original posterior. \cor{Notably, this approach also allows us to forecast ``how many instances of the studied events we would need to observe in order to claim a Proca-star detection'' respectively using frameworks that exploit and ignore boson-mass coincidences.}


Figure~\ref{fig:forecast_n} shows the $90\%$ credible intervals for the Bayesian evidence $p(\{d_i\}|n)$ for the number of bosons $n$ as a function of the number $N_{\rm obs}$ of repeated observations of a given set of events. In particular, the left panel ignores \cor{simulated events similar to S200114} while the right one includes them in the event sets. We respectively represent in red and green the natural logarithm of the ratio of the evidences for $n=1,2$ models {\it v.s.} the $n=0$ model, together with the $n=2$ {\it v.s.} $n=1$ ratio in magenta. In addition, we show in blue the result for the $n\neq0$ {\it v.s.} $n=0$, where the $n\neq0$ model ignores mass consistencies across events. The left panel shows results corresponding to $N_{\rm obs}$ instances of the 3-event sent ignoring S200114, while the right panel includes it. 

The blue contour of the left panel clearly shows that, if mass-consistencies are ignored, the $n\neq0$ model is progressively discarded as the number of observations increases. On the contrary, the green and red contours show that if mass-consistencies are taken into account, evidence for the $n\neq0$ models are progressively favoured. In particular, conclusive evidence is obtained for the $n=1$ model, with $\log {\cal B}^{n=1}_{n=0} > 5$ at the $90\%$ level after $N_{\rm obs}\simeq 8$ observations of the event set. The magenta contour shows that the preference for the $n=1$ over the $n=2$ one progressively increases as events accumulate. However, given the properties of the events involved (very similar masses and non very accurate mass estimates), obtaining conclusive preference for $n=1$ would require a very large number of repeated observations. In fact, we have checked that the $n=2$ model cannot be conclusively ruled out even if $N_{\rm obs}=50$.

The progressive preference for the $n = 1$ model is accompanied by progressively tighter estimates of the boson mass, whose $90\%$ credible interval is shown in the left panel of Fig.~\ref{fig:forecast_mass} (blue). The green and yellow contours of the same panel denote the boson mass estimates obtained with the $n=2$ model. As somewhat expected, such estimates have large uncertainties that always include the GW190521 median mass within the $90\%$ credible interval. This is due to the accuracy of the mass estimate being below the allowed minimum difference $\delta \mu_B$ between different masses. In other words, one cannot distinguish between a single boson model and a model with two bosons with extremely similar masses that differ by less than our maximum allowed resolution $\delta \mu_B$. 

The right panels of Figs.~\ref{fig:forecast_n} and~\ref{fig:forecast_mass} shows the same results discussed above but for the case where S200114 is included in the analysis. The only exception is that the magenta contour in the right panel of Fig.~\ref{fig:forecast_n} now corresponds to the ratio of the evidences between the $n=2$ and $n=1$ models. \cor{The inclusion of events similar to S200114 in the event sets has several implications}. First, conclusive evidence for $n\neq0$ can be obtained now after $\simeq 8$ repeated observations even if mass-coincidences are ignored due to the higher preference of S200114 for the PSM model. Second, such evidence is obtained after only $\simeq 4$ repeated observations if mass-coincidences are included using the $n=2$ model. Finally, conclusive evidence for the $n=2$ {\it v.s.} the $n=1$ model is obtained after $\simeq 9$ repeated observations. 

In terms of boson-mass estimations, the right panel of Fig.~\ref{fig:forecast_mass} shows that the two masses considered in the $n=2$ model progressively converge to the median values of S200114 and GW190521, with the former being more accurately estimated right from the start as a consequence of its larger preference for the PSM model. Additionally, we note that, given its restriction to a single boson, the $n=1$ model clearly discards GW190521 and similar simulated events as a PSMs, ignoring the corresponding boson mass estimates; opting instead to classify S200114 as a PSM, as somewhat expected.

\section{Discussion}

Evidence for a given type of GW source can be obtained either through strong evidence coming from a single a observation or through the accumulation of mild evidence across different events. GW sources like boson stars or neutron stars are described by characteristic parameters, respectively the underlying boson mass and the tidal deformability, which shall \cor{have} common \cor{value} across the whole source population. 

Focusing on the case of a particular class of boson stars known as Proca stars, we have shown how \cor{parameter} consistencies across events (\cor{in our case, the boson masss}) can be exploited to detect underlying populations of Proca stars and their properties even if no individual event can be conclusively identified as such. In particular, our formalism allows to progressively constrain the number of boson-star families and their characteristic masses as observations accumulate. The same formalism could be easily adapted to, for instance, constrain the number of families of compact objects populating the Universe characterised by {\it e.g.} their characteristic tidal deformabilities or the number and type of possible equations of state characterising neutron-stars and alternative compact objects.

Applied to the set of events we analyse, we conclude that, these are most consistent with all being black-hole mergers, {\it i.e.} we do not need to invoke any number of ultralight bosons to explain then. We note, however, that a number $n = 1$ and $n=2$ ultralight bosons are not conclusively discarded. Moreover, we show that due to the consistency of the boson-mass estimates across these events, \cor{our framework would make a value of $n=1$ be conclusively preferred after 5 to 9 observations of events with similar characteristics}. We also show that such evidence would not be achieved if ignoring the mentioned mass consistencies, which underscores the power of our formalism.

Consistently with previous work, we have checked that our results are heavily influenced by the intrinsic weakness of the Proca-star merger configurations with which the GW events have been compared, which imposes a strong penalty on the model. \cor{The impact of this ingredient is reflected in the fact that, if removing it, the hypothesis} that all the studied events are black hole mergers ({\it i.e.} $n=0$) is mildly rejected with Bayes Factors of $5$($40$) when the event S200114 is considered (discarded) as a true GW signal.

We note that our results are also heavily \cor{driven} by the limited number of currently available Proca-star merger simulations and, even more importantly, \cor{by} the restriction of these \cor{simulations} to almost ``toy models'' restricted to head-on mergers\cor{.} First, such configurations are astrophysically very unlikely. Second, as mentioned before, these lead to very weak signals that require the source to be way closer than a black-hole merger producing the same signal, which imposes a heavy statistical penalty. Third, the \cor{extremely short duration} of the these signals limits our study to very short GW signals from the heaviest black-hole merger detections, preventing the analysis of lighter systems with visible inspirals. Current work, {\it e.g.} Ref.~\cite{Palloni_Nico_ecc}, is ongoing towards simulating less eccentric configurations. Finally, the 769 numerical simulations to which our events were compared omit the impact of varying the relative phase of the fields \cor{of} the stars at merger. While in the simulations used in our input studies the relative phase of the stars is always fixed to zero, Ref.~\cite{Nico_phases} showed that varying this can have a dramatic impact in the signal. The future development of extended catalogs of numerical simulations and continuous surrogate models~\cite{Luna2024}, together with the increased sensitivity of the existing gravitational-wave detector network and the advent of next generation detectors such as LISA, Cosmic Explorer and Einstein Telescope~\cite{LISA,US_CE,Hild2011_ET,ET_Science} shall reveal or rule out the existence of compact exotic objects and, in particular, that of Proca Stars.\\

\section*{Acknowledgements }
ALM is supported by the Spanish Agencia Estatal de Investigaci\'{o}n through the grant PRE2022-102569, funded by MCIN/AEI/10.13039/ 501100011033 and the FSE+. This work has received financial support from Mar\'{ı}a de
Maeztu grant CEX2023-001318-M funded by MICIU/AEI/10.13039/501100011033, and from the Xunta de Galicia (CIGUS Network of Research Centres) and the European Union.  JCB is supported by the Ramon y Cajal Fellowship RYC2022-036203-I, by the Grant PID2024-160643NB-I00 of the Spanish Ministry of Science, Innovation and Universities; and by the Grant ED431F 2025/04 of the Galician Conselleria de Educaci\'on, Ciencia, Universidades e Formaci\'on Profesional. IGFAE is supported by the Ayuda Maria de Maeztu CEX2023-001318-M funded by MICIU/AEI /10.13039/501100011033. We acknowledge further support from the European Horizon Europe staff exchange (SE) programme HORIZON-MSCA2021-SE-01 Grant No. NewFunFiCO-101086251. SHWL acknowledges support by grants from the Research Grants Council of Hong Kong (Project No.~CUHK~14304622 and 14307923), and the Direct Grant for Research from the Research Committee of The Chinese University of Hong Kong. The authors acknowledge computational resources provided by the CIT cluster of the LIGO Laboratory and supported by National Science Foundation Grants PHY-0757058 and PHY0823459; and the support of the NSF CIT cluster for the provision of computational resources for our parameter inference runs. This material is based upon work supported by NSF's LIGO Laboratory which is a major facility fully funded by the National Science Foundation. This manuscript has LIGO DCC number P2500136. 


\bibliography{boson_pop.bib,psi4_observation,psi4_formalism}

\end{document}